\documentclass[sigplan,10pt]{acmart}

\usepackage{booktabs}
\usepackage{subcaption}
\usepackage{graphicx}
\usepackage{listings}            

\begin{document}

\title{A Serverless Publish/Subscribe System}

\titlenote{This is the comprehensive version of the work published at Middleware '17: Proceedings of the 18th ACM/IFIP/USENIX Middleware Conference: Posters and Demos, https://doi.org/10.1145/3155016.3155024~\cite{nasirifard}.}

\author{Pezhman Nasirifard}
\affiliation{
  \institution{Technical University of Munich}
   \country{Germany}
}
\email{p.nasirifard@tum.de}

\author{Hans-Arno Jacobsen}
\affiliation{
  \institution{University of Toronto}
  \country{Canada}
}

\renewcommand{\shortauthors}{P. Nasirifard et al.}

\begin{abstract}
Operating a scalable and reliable server application, such as publish/subscribe (pub/sub) systems, requires tremendous development efforts and resources. The emerging serverless paradigm simplifies the development and deployment of highly available applications by delegating most operational concerns to the cloud providers. The serverless paradigm describes a programming model where the developers break the application downs into smaller microservices that run on the cloud in response to events. This paper proposes designing a serverless pub/sub system based on the IBM Bluemix cloud platform. Our pub/sub system performs topic-based, content-based, and function-based matchings. The function-based matching is a novel matching approach where the subscribers can define a highly customizable subscription function that the broker applies to the publications in the cloud. The evaluations of the designed system verify the practicality of the designed system. However, the vendor-specific constraints of the IBM Bluemix resources are a bottleneck to the scalability of the broker.
\end{abstract}

\settopmatter{printacmref=false}




\settopmatter{printacmref=false}
\setcopyright{none}
\renewcommand\footnotetextcopyrightpermission[1]{}
\pagestyle{plain}
\settopmatter{printfolios=true}

\maketitle

%%%%%%%%%%%%%%%%%%%%%%%%%%%%%%%START%%%%%%%%%%%%%%%%%%
\section{Introduction} 
\label{sec:introduction}

Distributed messaging applications such as publish/subscribe (pub/sub) systems play a critical role in operating distributed applications. A pub/sub system relies on several middleware applications to receive the publications and deliver the matching publications to the subscribers \cite{many_faces_pub_sub}. However, developing and operating a large-scale distributed pub/sub system raises several concerns with scalability, reliability, availability, and fault tolerance. The development and administrators teams often have to foresee and manage \cite{dist_book}. These additional mandatory requirements add up to the complexity of the application, which increases the required time, effort, engineering knowledge, and cost for creating the system.

Recently, serverless computing has emerged as a new paradigm in cloud computing. Serverless computing simplifies the application development, distribution, and deployment cycle by delegating most of the operational concerns from development teams to the cloud providers \cite{serverless_compueting_trends}. Serverless computing, also known the Function-as-a-Service (FaaS), represents a programming model where the developers decompose the applications into microservices, also known as functions, and deploy the functions to the cloud. Next, the cloud providers are responsible for executing the functions in response to events and performing the automatic monitoring and maintenance of the application's health, resources, and performance. Serverless computing opens a new market for cloud providers. Also, more importantly, serverless resources liberate the developers from most of the traditional maintenance effort of server applications so that the developers can focus more on creating the application. Currently, all primary cloud vendors such as Amazon Web Services \cite{aws}, IBM Bluemix \cite{ibm_bluemix}, Microsoft Azure \cite{azure} and Google Cloud Platform \cite{google} provide serverless platforms.

Despite the apparent benefits of serverless computing, developing serverless applications are not the simple task of deploying the source code of the legacy applications to the serverless platform. The developers require a different approach to implementing the serverless applications to address the serverless platform requirements and limitations. One primary limitation of developing serverless applications is the short-lived stateless nature of serverless microservices \cite{occupy_cloud}. Serverless platforms do not persist the state of the functions across multiple executions to improve stability. However, since the cloud providers integrate the serverless platforms into the cloud ecosystem, the serverless platforms can use the storage services offered by cloud providers to persist the state of the serverless applications. 

In this paper, we investigate the applicability of the serverless paradigm to scalable stateful distributed pub/sub systems. We propose a design for a serverless pub/sub system that performs topic-based, content-based, and function-based matching. The function-based matching is a novel matching approach that enables subscribers to create customized and complicated matching logic. We also offer an implementation of our design based on IBM Bluemix. We use IBM Cloud Functions \cite{openwhisk} to receive the publications and subscriptions and to perform the matching methodologies. We also use the IBM Cloudant database \cite{cloudant} to persist the application's state, and the IBM Watson IoT platform \cite{watson} to deliver the publications to the subscribers. Furthermore, we conduct experiments to measure our serverless approach's latency, cost, and performance. 

We organize the remainder of the paper as follows. In Section \ref{sec:background}, we provide an overview of the IBM Bluemix technologies. In Section \ref{sec:related_work}, we review existing serverless approaches and applications. Next, in Section \ref{sec:approach}, we describe the designed pub/sub system and we discuss the evaluations and limitations of our implementation in Section \ref{sec:evaluation}. Finally, we present the concluding remarks and future work in Section \ref{sec:conclusion}. 

%%%%%%%%%%%%%%%%%%%%%%%%%%%%%%%%%%%%%%%%%%%%%%%%%%%%
\section{Background}
 \label{sec:background}

We develop and deploy the pub/sub system on IBM Bluemix \cite{ibm_bluemix}, a cloud platform operated by IBM that provides a large variety of cloud resources and services. We make use of three Bluemix technologies, including IBM Cloud Functions \cite{openwhisk, nasirifard_2} as a serverless platform, the Cloudant NoSQL database \cite{cloudant}, and IBM Watson IoT \cite{watson} as a communication channel.

IBM Cloud Functions is a serverless FaaS platform based on Apache OpenWhisk \cite{apache_openw}. Cloud Functions is hosted and operated by Bluemix and provides an event-based serverless programming model where a stateless function, also known as an \textit{Action}, is executed either upon receiving an event, called a \textit{Trigger} or through the provided REST API. The actions accept a dictionary object, such as a JSON object, as an input and return a dictionary object as an output. The actions contain a block of code in any of the supported languages, including but not limited to JavaScript and Python, and can be chained together to create a more complex action known as a \textit{Sequence}. Triggers invoke the actions according to the predefined association with an action, also known as a \textit{Rule}. Moreover, the triggers can be caused by many external and internal event sources, such as a database change or a file upload to the storage. The developers can bundle a set of related actions together as a \textit{Package} which other actions can use. Finally, Cloud Functions is responsible for automatically managing all the necessary resources for deploying and executing actions and automatically scaling up and down according to the demand. 

IBM Bluemix provides the Cloudant NoSQL database, a scalable document-oriented Database-as-a-Service (DBaaS), to store and query data in JSON format. Cloudant DB is accessible in the Cloud Functions ecosystem for actions to store and retrieve data. The developers can also define triggers that invoke actions in response to the data changes on the Cloudant DB.

Finally, IBM Watson IoT is a robust platform for realizing IoT infrastructures offering a broad range of features from natural language processing to risk and security management. However, most of the features are out of the scope of this paper. We exploit Watson IoT to register the subscribers as IoT devices and to establish a secure real-time bi-directional channel for delivering the publications to subscribers.

%%%%%%%%%%%%%%%%%%%%%%%%%%%%%%%%%%%%%%%%%%%%%%%%%%%%
\section{Related Work} 
\label{sec:related_work}

A few studies discuss the applicability of serverless computing to different use cases and address the serverless paradigm's open problems. Serverless computing is a practical tool for executing many use cases, from video processing to power grid operations management~\cite{video_processing_with_thousands, nasirifard_3, nasirifard_5, nasirifard_4}. Fouladi et al.~\cite{video_processing_with_thousands} present a serverless video processing system for executing thousands of cloud functions in parallel to edit, transform and encode videos. Their evaluation shows that by taking advantage of serverless architecture for running computationally heavy tasks in parallel, they decrease the latency of video manipulation operations. Similar to their approach, we benefit from serverless resources to run a large number of functions in parallel for performing publications matching and delivering the publications. Moreover, Yan et al. \cite{serverless_chatbot} present a prototype of a chatbot based on a serverless platform, and Ast et al. \cite{self_contained} demonstrate a serverless approach for encapsulating and deploying web components. Both works show that using serverless resources reduces; they decreased the necessary effort for developing applications and maintaining the scalability and extensibility of the application. 

The current available serverless platforms are stateless, and enabling the serverless paradigm with a persistent state is yet an open problem \cite{serverless_compueting_trends}. Hendrickson et al. \cite{serverless_with_openlambda} proposes an open source platform for building applications based on serverless computing, called \textit{OpenLambda}. In their approach, they suggest using cookies to persist the state of the sessions. They also recommend a better integration of managed database services. Additionally, Spillne \cite{snafu} proposes a design of a modular system to maintain and execute serverless microservices, called \textit{Stafu}. Spillne also suggests integrating stateful services such as key-value stores and file storage to persist the services' state. Similarly, we store the necessary data in the provided Database-as-a-Service to compensate for the lack of state.

Since serverless computing is a somewhat new concept, the scientific work on integrating the serverless paradigm with pub/sub systems is rare \cite{serverless_programming_short}. Nevertheless, Nasirifard et al. \cite{nasirifard} demonstrate a serverless pub/sub broker that performs content-based and topic-based matching. We extend their work by improving their matching approaches by caching the serverless functions' subscriptions and proposing a novel function-based matching scheme. Furthermore, we offer an evaluation of our pub/sub system performance and latency which was missing from the previous works.  

%%%%%%%%%%%%%%%%%%%%%%%%%%%%%%%%%%%%%%%%%%%%%%%%%
\section{Approach} 
\label{sec:approach}

In this section, we present the architecture of the designed pub/sub system. First, we describe the realized use cases, then explain the implementation in detail.

\subsection{Use Case Model} 
\label{subssec:usecase_model}

A pub/sub system consists of three primary parts: subscribers, publishers, and brokers. The subscribers express their interest in the specific data type to the broker; the publishers broadcast publications containing some data to the broker. The brokers receive the publications and use a matching scheme to match the publication with the subscriptions and forward the matching publications to the filtered subscribers. The broker can support different matching schemes based on the type of publications and subscriptions.

\begin{figure}
\includegraphics[width=3in]{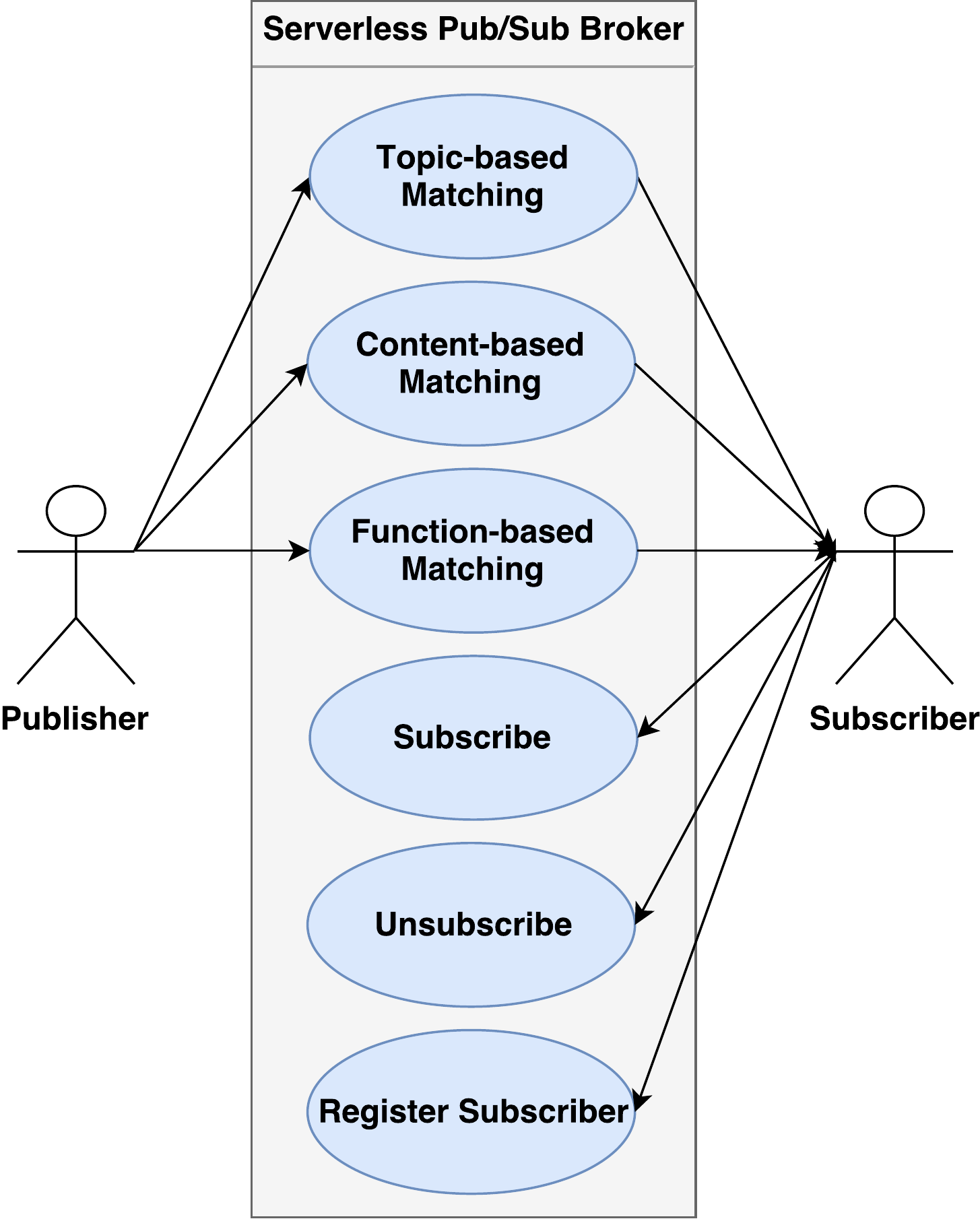}
\caption{Use case diagram of the serverless pub/sub system.}
\label{fig:use_case}
\end{figure}

We realize six primary use cases for our design, represented by the UML use case diagram of Figure \ref{fig:use_case}, including \textit{Register Subscriber}, \textit{Subscribe}, \textit{Unsubscribe}, \textit{Topic-based Matching}, \textit{Content-based Matching} and \textit{Function-based Matching}. The subscribers are required to register themselves before any interaction with the broker. The broker assigns a unique identifier to the new subscriber and stores the ID in the database. Afterward, the subscriber can create and submit subscriptions, which the broker receives and stores in the database. Additionally, the subscribers can remove the subscriptions from the broker. Meanwhile, the publisher creates publications, which, based on the matching scheme, have different structures, and then submits the publications to the broker. Immediately upon delivery to the broker, the broker performs the matching operation based on the chosen matching scheme and forwards the matching publications to the filtered subscribers. 

\subsection{The Serverless Broker Matching Schemes}
\label{subsec:matching_schemes}

As previously stated, the designed serverless system performs three matching schemes: the two widely used topic-based and content-based matchings and the novel function-based matching scheme.

\subsubsection{Topic-based Matching} 
\label{subsubsec:topic_based}

For topic-based matching, the publishers create publications with a set of topics or subjects assigned to the publication, which the broker receives and forwards to the subscribers that previously subscribed to the topics. For topic-based matching, the publications should be formatted like $(publicationData, [topic_1, topic_2, ..., topic_n])$ where the subscribers can create subscriptions like \\$[topic_1, topic_2, ..., topic_n]$. The broker forwards the publication to any subscriptions that contain at least one of the assigned topics to the publication. 

\subsubsection{Content-based Matching} 
\label{subsubsec:content_based}

Content-based matching offers a more expressive matching approach where the subscriptions are defined and matched against the publication's content. In more detail, the publications contain an arbitrary number of key-value tuples defining the properties of the publication, and the subscriptions contain some constraints for filtering the publications in the form of key-value pairs of properties and basic comparison operators (=, <, <=, >, >=) \cite{many_faces_pub_sub}. In our system, the publication should be formatted like $(publicationData, [key_ 1: value_1, key_2 : value_2, ..., key_n : value_n])$ and the subscription should be created based on the $[key_1 = value_1, key_2 > value_2, ..., key_n <= value_n]$. The broker forwards the publication when the assigned properties fully satisfy the subscription constraints. 

\subsubsection{Function-based Matching} 
\label{subsubsec:function_based}

We introduce a novel and extremely expressive matching approach that we name function-based matching, which we realize based on the flexible execution environment of the serverless platforms. The subscribers create and submit a matching function as a subscription for a function-based matching scheme. The matching function can contain a Turing-complete code implemented in any supported programming language of the serverless platform, which the serverless broker applies to the publication's content. In our designed system, we format the publication like $(publicationData, matchingFunctionType)$, where the matching function type is a simple tag explaining the category of the matching function. We define the subscription as $(matchingFunctionType, functionCode)$. The broker applies the matching function to publications of the same type. If the publication's content satisfies the function's logic, the broker forwards the matching publication to the filtered subscribers. We should emphasize that the matching function can contain any arbitrary logic, from simple calculations based on the publication's content to performing complex image recognition methods. 

\subsection{The Broker's Implementation}
\label{subsec:prototype_impel}

\begin{figure*}
\includegraphics[scale=0.55]{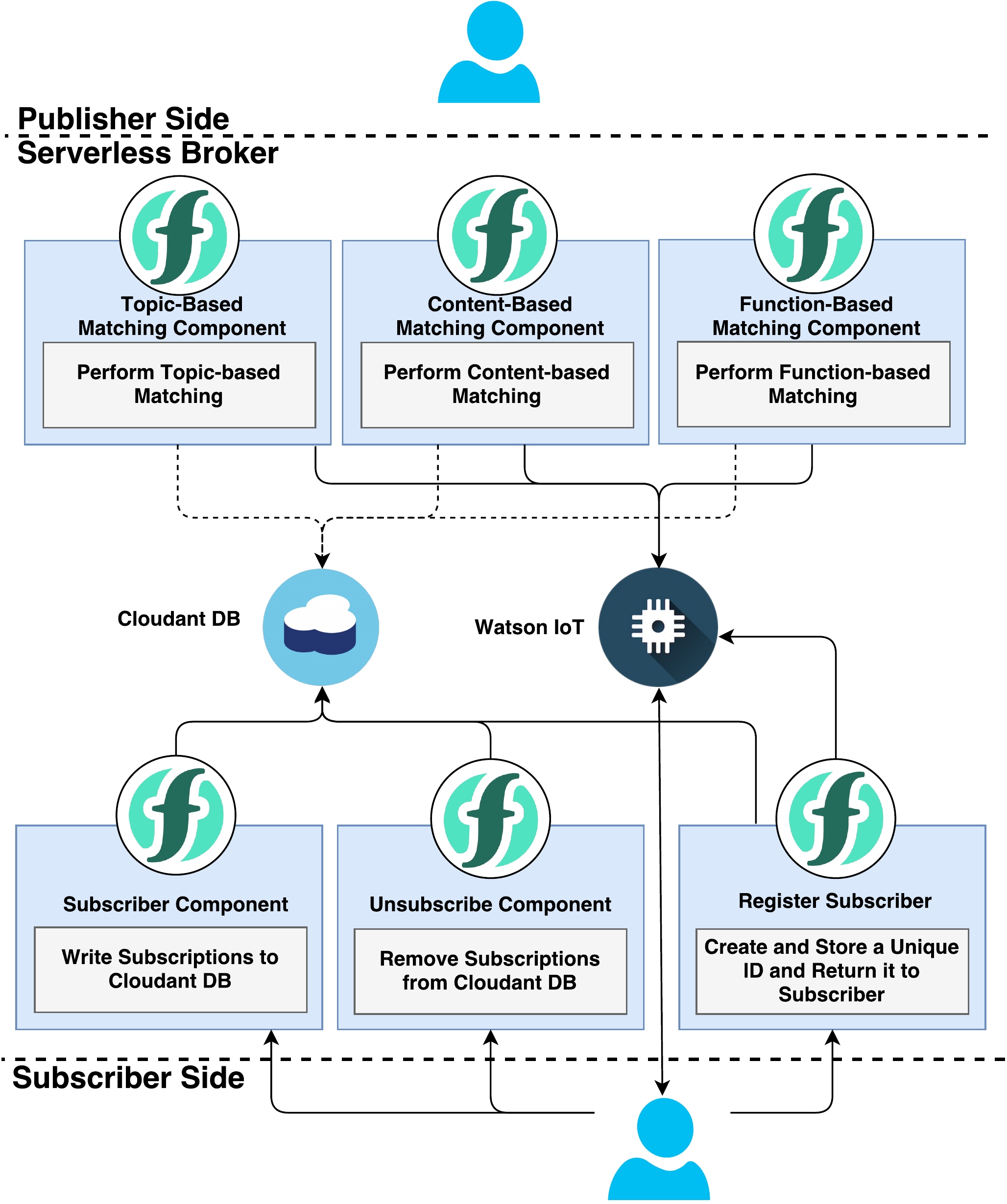}
\caption{The system decomposition of the serverless pub/sub system.}
\label{fig:serverless_workflow}
\end{figure*}

We decompose the realized use cases into microservices which we develop as actions on top of the serverless platform, as Figure \ref{fig:serverless_workflow} displays. Each use case consists of multiple actions which are executed upon receiving requests from publishers or subscribers. Furthermore, the actions use Cloudant DB to persist the state subscriptions and Watson IoT to deliver the matching publications to subscribers. In the following,  we explain the detailed implementation of each use case. 

\subsubsection{Register Subscriber} 
\label{subsubsec:register_sub}

The subscriber client executes this use case in two phases. First, the subscriber client sends an HTTP GET request to the action, which generates a unique identifier (ID) for the new subscriber, inserts the new ID into the Cloudant DB, and then returns the ID to the subscriber client. The subscriber client receives the ID and submits the ID to the Watson IoT platform through an HTTP POST to register the client as an IoT device. 

\subsubsection{Subscribe and Unsubscribe} 
\label{subsubsec:sub_topics}

\begin{figure}
\includegraphics[width=.8\columnwidth]{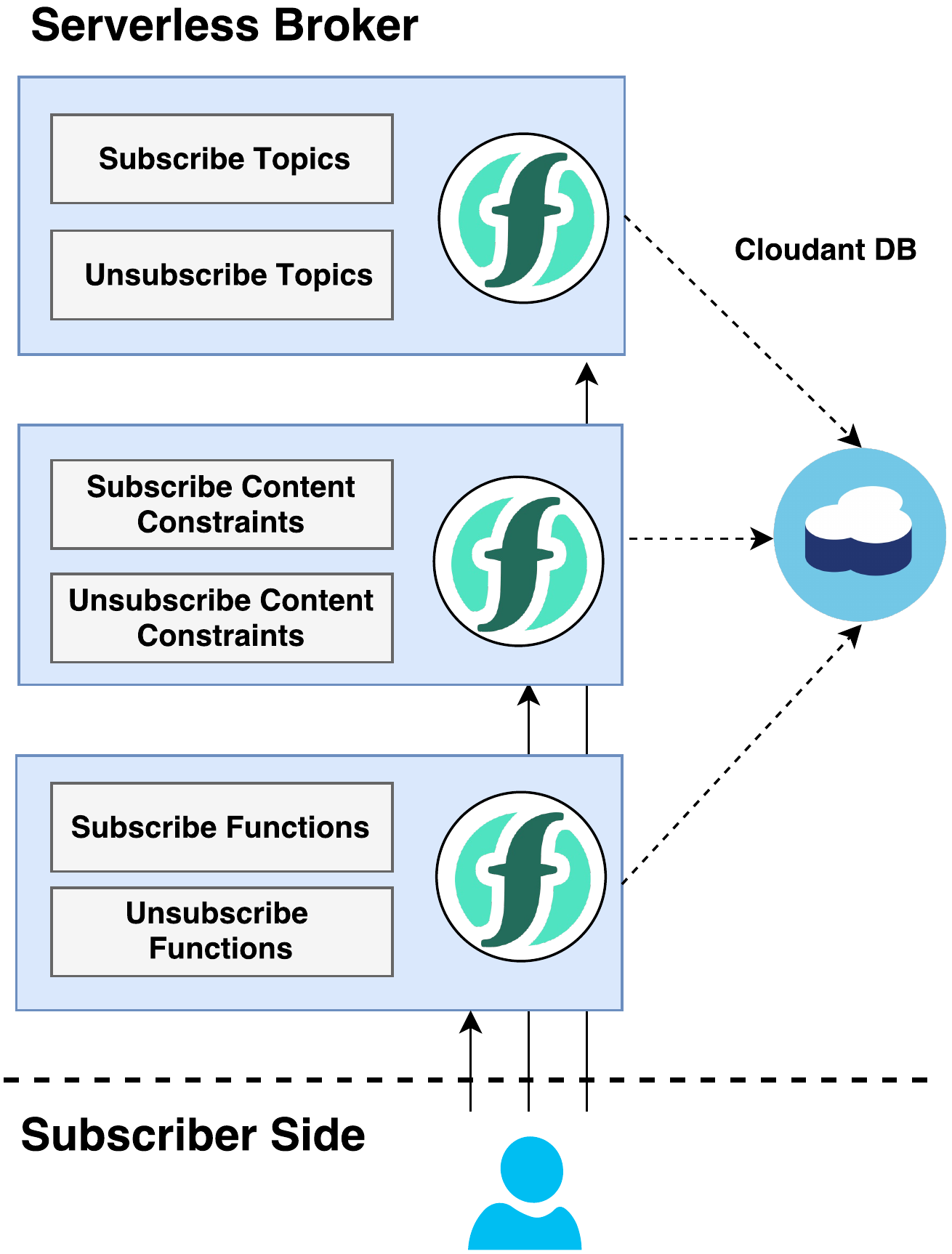}
\caption{Decomposition of subscription and unsubscription use cases.}
\label{fig:sub_unsub}
\end{figure}

The implementation of these two use cases is almost identical, so we describe them together. However, the procedures for creating and removing subscriptions of different matching schemes are different, as Figure \ref{fig:sub_unsub} displays, although we previously grouped the use cases for the sake of simplicity. 

When creating or removing a topic-based related subscription, the subscriber submits a list of topics with the subscriber ID like $(subscriberID, [topic_1, topic_2, ..., topic_n])$ to the serverless action through an HTTP POST request. For a subscription request, the action redundantly stores the received subscriber ID and topics in two different data sets in Cloudant DB; each topic is stored in a separate document on the database, which contains a list of the topic's subscribers. Also, each subscriber owns an individual document on the Cloudant DB containing all the subscribed topics. The main reason for this redundancy is improving the performance of Cloudant DB, as executing a lookup based on the document's ID is more efficient than running queries on the content of the documents to fetch the subscriptions. For an unsubscription request, the action removes the subscriber ID from the topic's document and the topics from the subscriber's document.

For creating a content-based related subscription, the subscriber submits a list of key-value properties with the subscriber ID like $(subscriberID, [key_1 > value_1, key_2 >= $ \\ $ value_2, ..., key_3 = value_3])$ to the serverless action through an HTTP POST request. Similar to topic-based matching, the action stores the received subscription in two sets of datasets in Cloudant DB. First, the action splits the key-value list into key and value pairs where each key is stored in a separate document on the database, which contains a list of the IDs of all subscribers who are subscribed to the key and the subscription property's value. Furthermore, each subscriber owns an individual document on the Cloudant DB containing all the subscribed key-value properties. We should note that our system currently supports one set of content-based subscriptions for each subscriber and a new subscription overwrite the previous subscription. For an unsubscription request, the action removes the subscriber ID from the stored key-value subscription properties document, and the action also removes the subscription from the subscriber's document.

Finally, when creating a function-based related subscription, the subscriber submits a subscription to the serverless action via an HTTP POST request that contains the subscriber ID, the matching function's type, and the source code of the matching function encoded as a string. The subscription has the format of $(subscriberID, matchingFunctionType,$ $ functionCode)$. Similar to the previous subscriptions, the subscription is stored in two data sets in Cloudant DB. First, each function type is stored in a separate document on the database, containing a list of the IDs of all subscribers subscribed to the same matching function and the subscription's function code. Furthermore, each subscriber owns an individual document on the Cloudant DB containing all the function-based subscriptions. Each subscriber can have unlimited subscriptions with a different function type. However, there can only exist one function code per function type; submitting a new subscription with an existing function type overwrites the previous subscription with the same function type. For removing a function-based subscription, the subscriber submits a request containing the subscriber ID and the function type, like $(subscriberID, matchingFunctionType)$, where the serverless action removes the subscriptions from respective datasets.

\subsubsection{Topic-based Matching} 
\label{subsubsec:topic_based_matching}

\begin{figure}
\includegraphics[width=.8\columnwidth]{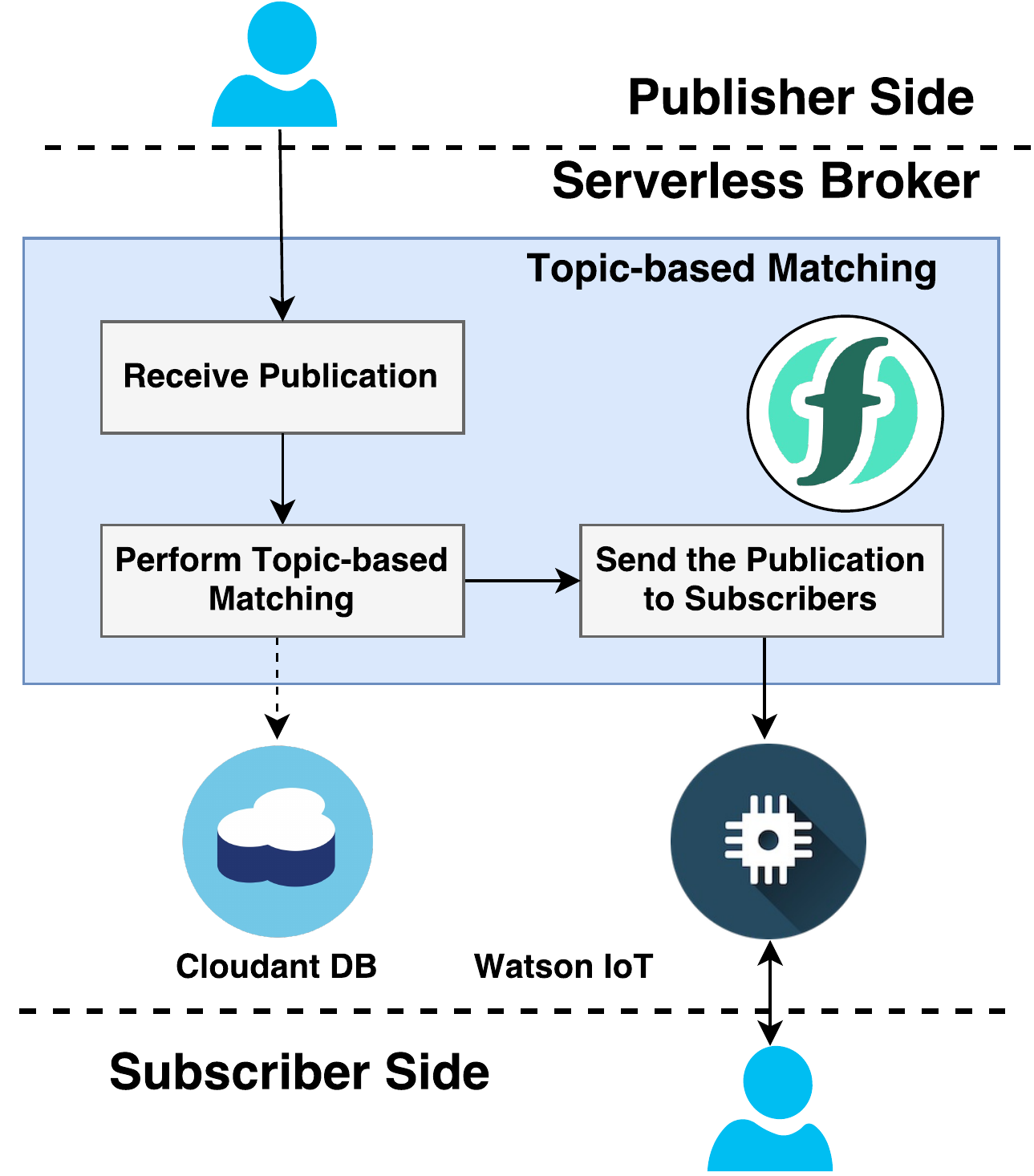}
\caption{Topic-based matching actions and workflow.}
\label{fig:topic_match}
\end{figure}

To perform a topic-based matching on the publications, the serverless broker follows a procedure consisting of three serverless actions, as Figure \ref{fig:topic_match} displays. An arbitrary number of publishers and subscribers can connect to the serverless broker hosted on the serverless platform. 

To submit a publication to the broker, the publisher creates and submits a publication like $(publicationData, [topic_1, $ $ topic_2, ..., topic_n])$ to the first action via an HTTP POST request. In this study, the publication data is limited to text strings for simplicity, but the published data on the system can be any JSON object containing complex data structures. Once the action receives the publication, the action splits the assigned topics into a list of topics. For each topic, in parallel, the action forwards the publication data and one of the topics to the second action. 

The second action performs topic-based matching. To improve the action's performance and reduce the overall latency when fetching the subscription from Cloudant DB, the action uses a local cache. A local cache is a key-value dictionary stored inside the action, where the key is the topic, and the value is the list of subscriber IDs subscribed to the topic. First, the action looks for the topic in the cache; if it is a hit and the topic's subscribers have been updated within the past ten seconds, the action considers the list of subscribers to be the matching subscriptions. If it is a cache miss or the topic's subscribers are stale, then the action lookups at the topic in the Cloudant DB retrieve the topic's subscribers and store them in the cache. We should note that, in this work, we consider the data older than ten seconds to be stale. Finally, for each subscriber in the list, in parallel, the action passes the subscriber's ID, the publication data, and the topic to the third serverless action. 

Finally, the third action submits an HTTP POST request to the Watson IoT platform containing the subscriber ID, the publication data, the assigned topic, and a timestamp. Afterward, the Watson IoT platform forwards the message to the connected subscriber with the specified ID. 

\subsubsection{Content-based Matching} 
\label{subsubsec:content_based_matching}

\begin{figure}
\includegraphics[width=.8\columnwidth]{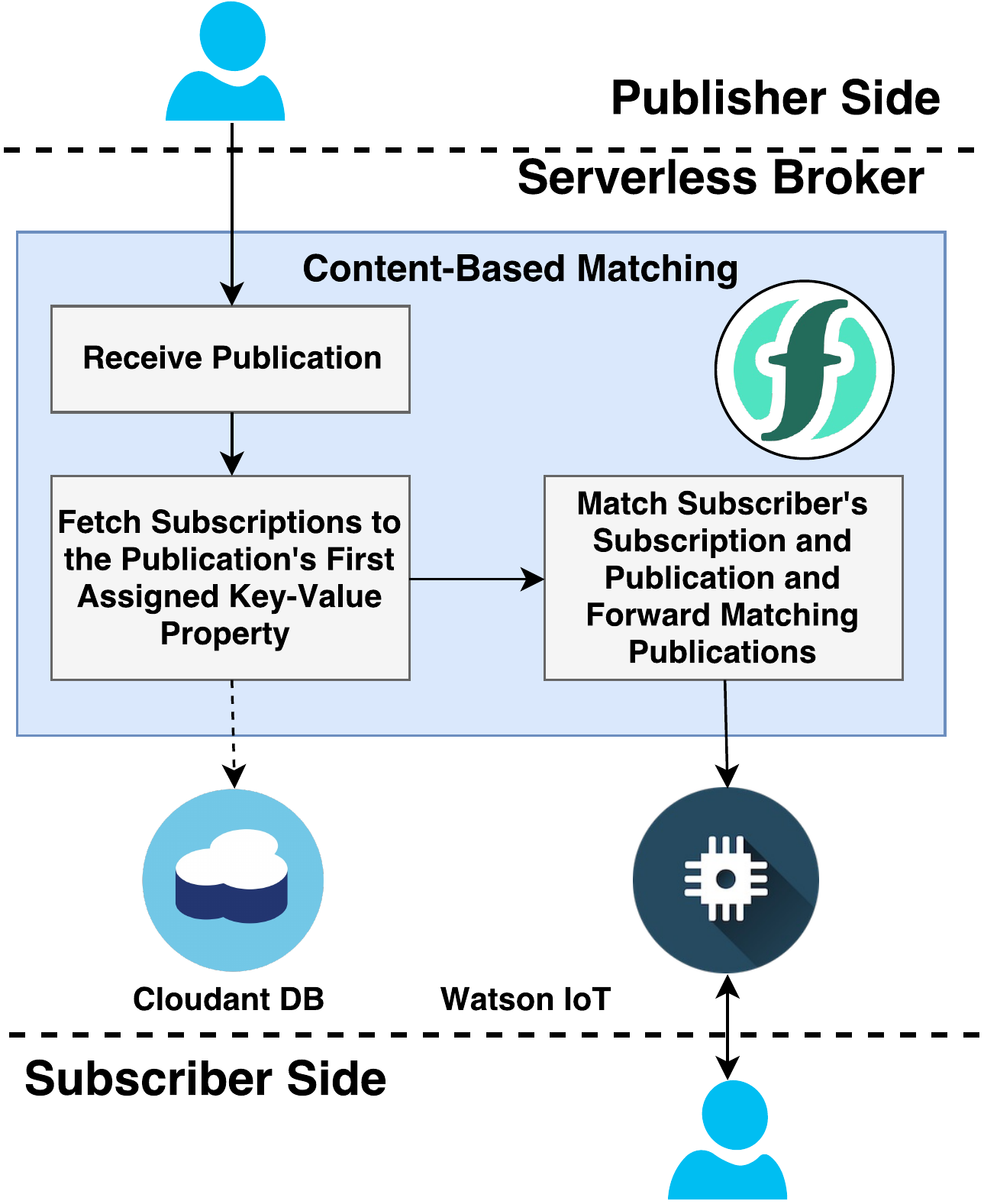}
\caption{Content-based matching actions and workflow.}
\label{fig:content_match}
\end{figure}

The serverless broker follows a workflow consisting of three serverless actions to perform content-based matching, as Figure \ref{fig:content_match} shows. The publisher creates and submits publication which adheres to a format like $(publicationData, [key_ 1: value_1, key_2 : value_2, ..., key_n : value_n])$. The first action receives the publication, splits the key-value properties, and sorts the pairs alphabetically based on the key. Then, the action forwards the publication data, the assigned key-value properties, and the first key in the sorted list to the second action. 

The second action receives input from the previous action. It uses a local cache similar to the topic-based approach to store the key of the key-value property and the subscription and the subscriber IDs of the subscribers. The key is a part of their content-based subscription. First, the action looks for the key in the cache; if it is a cache miss or the topic's subscribers are stale, then the action lookups for the key in the Cloudant DB to fetch the key's subscriptions and subscriber ID and put the result in the cache. Finally, for each subscription in the list, in parallel, the action passes the subscriber ID, the subscription, the publication data, and the key-value properties of the publication to the third serverless action. 

The third action receives the mentioned inputs and matches the subscription with the publication key-value properties. If the subscription and publication match, then the action forwards the publication data and the key-value properties of the publication, the subscriber ID, and a timestamp to the Watson IoT, which forwards the publication to the subscriber with the specified ID. 

\subsubsection{Function-based Matching} 
\label{subsubsec:function_based_matching}

\begin{figure}
\includegraphics[width=.8\columnwidth]{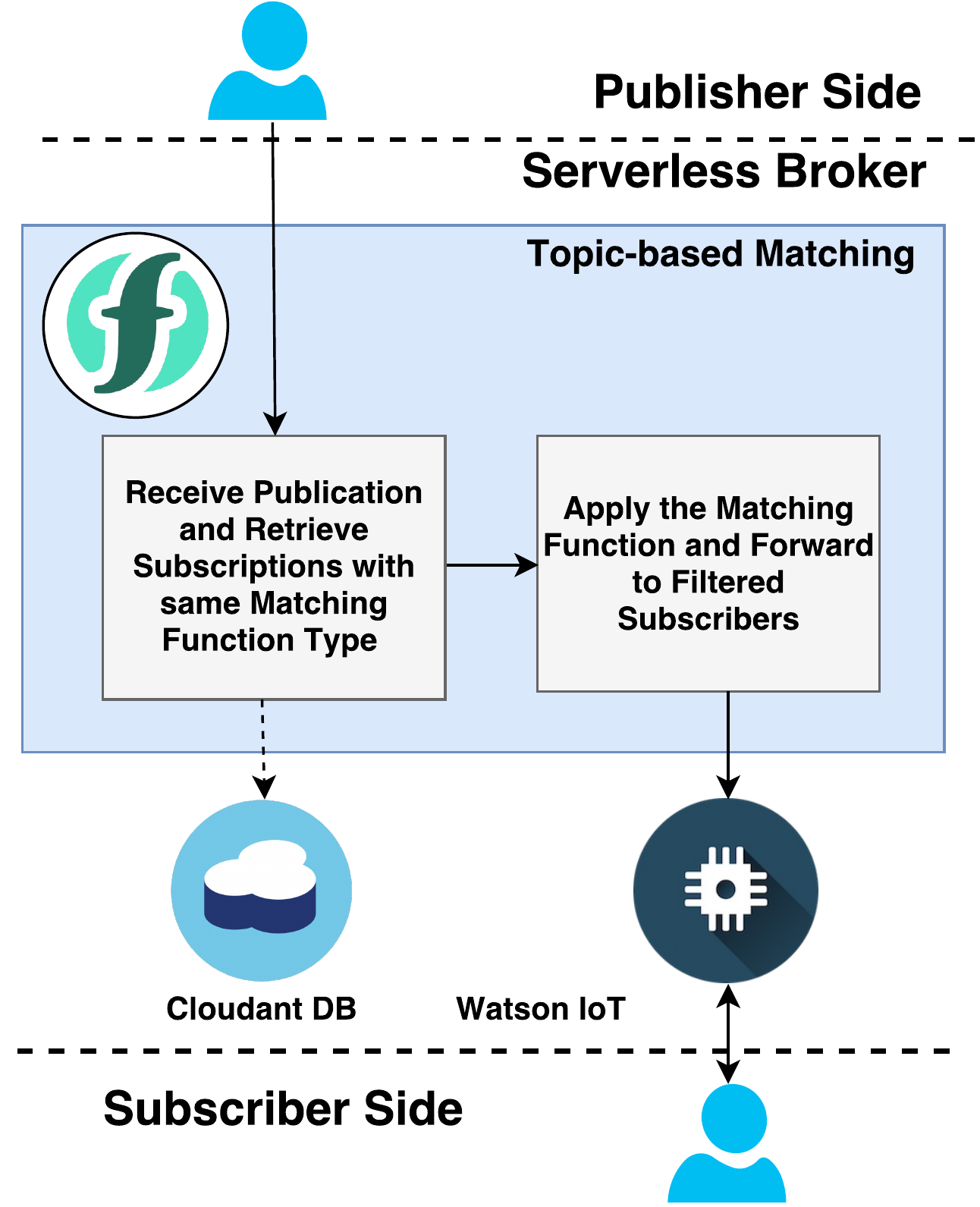}
\caption{Function-based matching actions and workflow.}
\label{fig:function_match}
\end{figure}

The function-based matching workflow consists of two actions, as Figure \ref{fig:function_match} displays. The publisher creates and submits publications like \\$(publicationData, matchingFunctionType)$. The first action uses a local cache consisting of a key-value dictionary where the key is the matching function type, the value is the list of subscriber IDs, and the function codes with a subscription with the same function type. First, the action looks for the publication's function type in the cache; if it is a cache miss or the cache is stale, then the action lookups for the publication function type to fetch the subscriber IDs and the function code and puts the results in the cache. Finally, for each subscriber ID in the list, in parallel, the action forwards the subscriber ID, the matching function type, the matching function code, and the publication data to the second action. 

Once the second action receives the inputs from the previous action, the action executes the matching function code, with the publication data passed into the function as input within the action's environment. Suppose the publication data satisfies the constraints of the matching function, meaning that the function returns a \textit{True} value. In that case, the action forwards the subscriber ID, the publication data, the function type, and a timestamp to the Watson IoT platform, which delivers the publication to the subscriber. \\

All the actions on IBM Cloud Functions are implemented in JavaScript, and we use a custom bash script to deploy the actions on the IBM Bluemix. The source code of the serverless broker is open-sourced and freely available under Apache License 2.0\footnote{\url{https://github.com/epezhman/i13BluemixBroker}}. We provide a cross-platform desktop application based on Node.js \textit{Electron Framework} \cite{electron} to execute the described functionalities of publisher and subscribers clients. The source code of the application is also available~\footnote{\url{https://github.com/epezhman/i13PubSub}}.

%%%%%%%%%%%%%%%%%%%%%%%%%%%%%%%%%%%%%%%%%%%%%%%%%
\section{Evaluation} 
\label{sec:evaluation}

To evaluate the performance and latency of our system, we conducted several experiments. The following section presents our experimental methodologies and the evaluation results. Finally, we discuss our findings and describe the limitations we faced during the development and evaluation of the system.

\subsection{Experimental Setup} 
\label{subsec:experiment_setup}

We designed a distributed experiment to measure the latency of the broker in response to an increasing number of subscribers and publications per second throughput. For each experiment, one publisher created and published an exact number of publications per second. On the subscriber side, an accurate number of subscribers are evenly distributed among four virtual machines and were connected to the broker simultaneously, e.g., for the experiment with 352 subscribers with 40 publications per second, each of the four virtual machines ran 88 subscribers, and each subscriber received in total 40 publications. 

To evaluate the topic-based scheme, we created publications with the format $(publicationData, [topic])$ where the size of publication data was 1 KB, and we assigned only one topic to the publication. For content-based matching, the used publication was formatted like $(publicationData, [key_1 : value_ 1, key_2 : value_2])$ where the size of publication data was also 1 KB, and the subscription had the format like $[key_1 = value_1, key_2 >= value_2]$. For the function-based experiments, we implemented a simple matching function as the subscription, which used a JavaScript-based natural language processing library called \textit{Compromise} \cite{compromise}, to extract a location from the publication data and verify if a condition for the extracted location holds, as Listing \ref{lst:match_code} illustrates. The publication has the format like $(publicationData, matchingFunctionType)$ where the publication data was the string "DEBS2018 will be held at the University of Waikato in New Zealand." which would result as \textit{True} when passed into the matching function. We should note that all the running subscribers in each experiment were subscribed to every submitted publication.

\begin{lstlisting}[caption={The matching function for the evaluation of function-based matching.},label=lst:match_code, frame=single, basicstyle=\footnotesize]  

function matchingPopulation(publication)
{ 
 const nlp = require('compromise');
 let populations = {'new zealand': 4693000, 
                   'germany': 8267000};
 let sentence = nlp(publication); 
 let places = sentence.places().out('array'); 
 return populations[places[0]] > 4000000;
}
\end{lstlisting}

At the end of each experiment, we measured the latency for each subscriber for receiving all of the publications. The objective of this experiment was to discover how the latency of the serverless broker changed as the number of publications throughput, and the number of subscribers increased. We repeated each experiment three times to minimize the effects of outliers and calculated the arithmetic mean of recorded results. 

The deployed actions on the IBM Cloud Functions were executed with a 256 MB memory limit runtime configuration and 60 seconds execution time limit. Furthermore, we were limited to 1200 simultaneous action execution and 9000 action invocation per minute on the IBM Cloud Functions. We chose the \textit{Standard Pay-As-You-Go} plan for Cloudant DB, which restricts the database interactions to 1000 lookups per second. We used the Watson IoT platform under \textit{Standard Pay-As-You-Go} plan. All the used IBM Bluemix technologies were running on the servers in Germany. Finally, the publisher client was running on a virtual machine, and subscribers were distributed evenly on four identical virtual machines. Each virtual machine used four virtual CPU cores and 9.77 GB of RAM, and 20 GB of SSD disk running an Ubuntu 16.04.

\subsection{Results and Findings} 
\label{subsec:results_findings}

\begin{figure*}[!t]
  \centering
    \begin{subfigure}{.45\textwidth}
  \centering
  \includegraphics[width=1\linewidth]{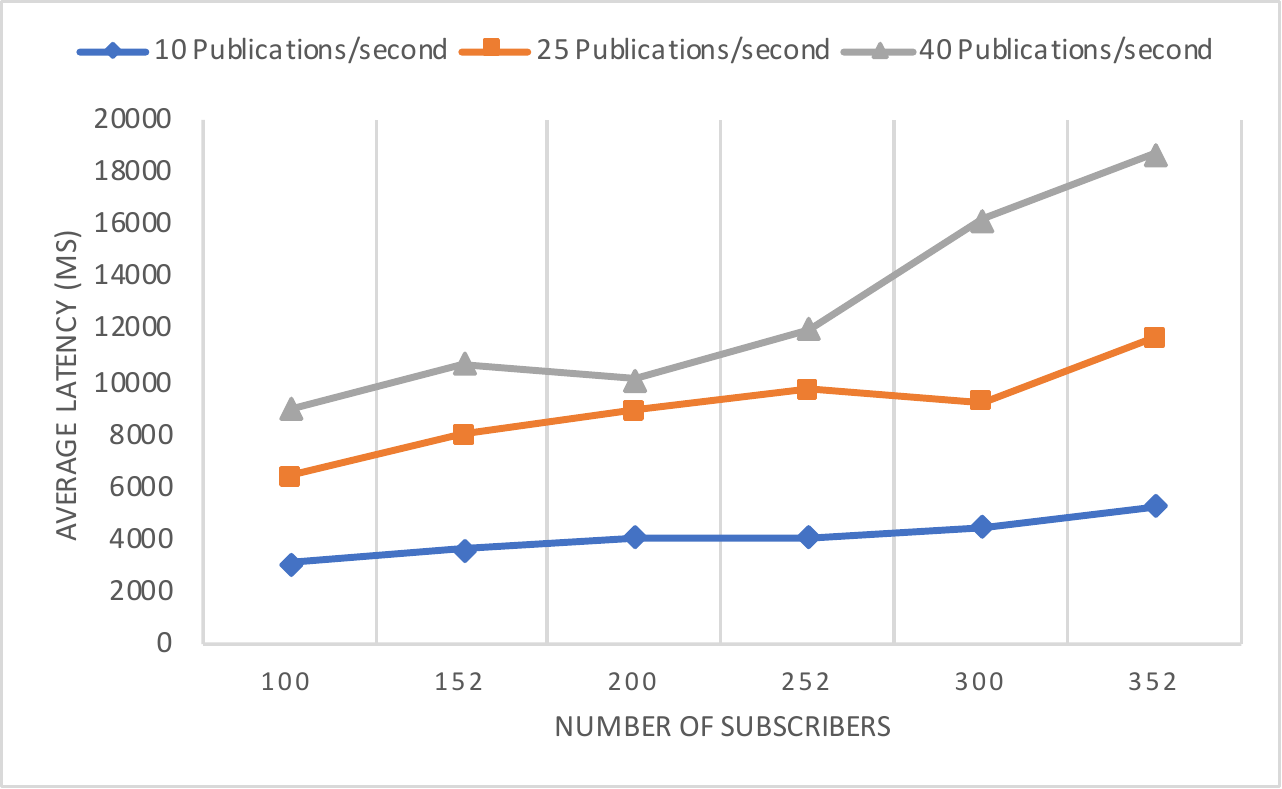}
  \caption{Topic-based matching. }
\end{subfigure}
\begin{subfigure}{.45\textwidth}
  \centering
 \includegraphics[width=1\linewidth]{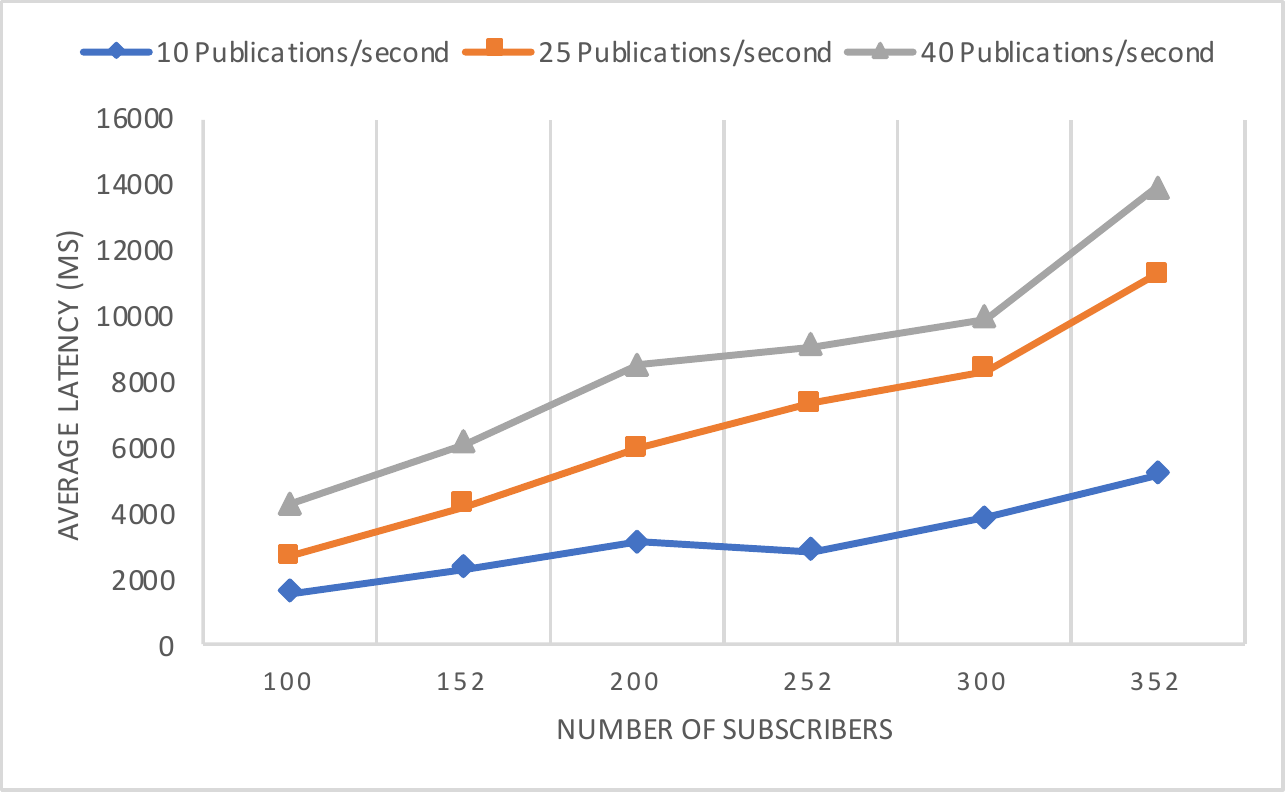}
  \caption{Content-based matching. }
\end{subfigure}
  \begin{subfigure}{.45\textwidth}
  \centering
  \includegraphics[width=1\linewidth]{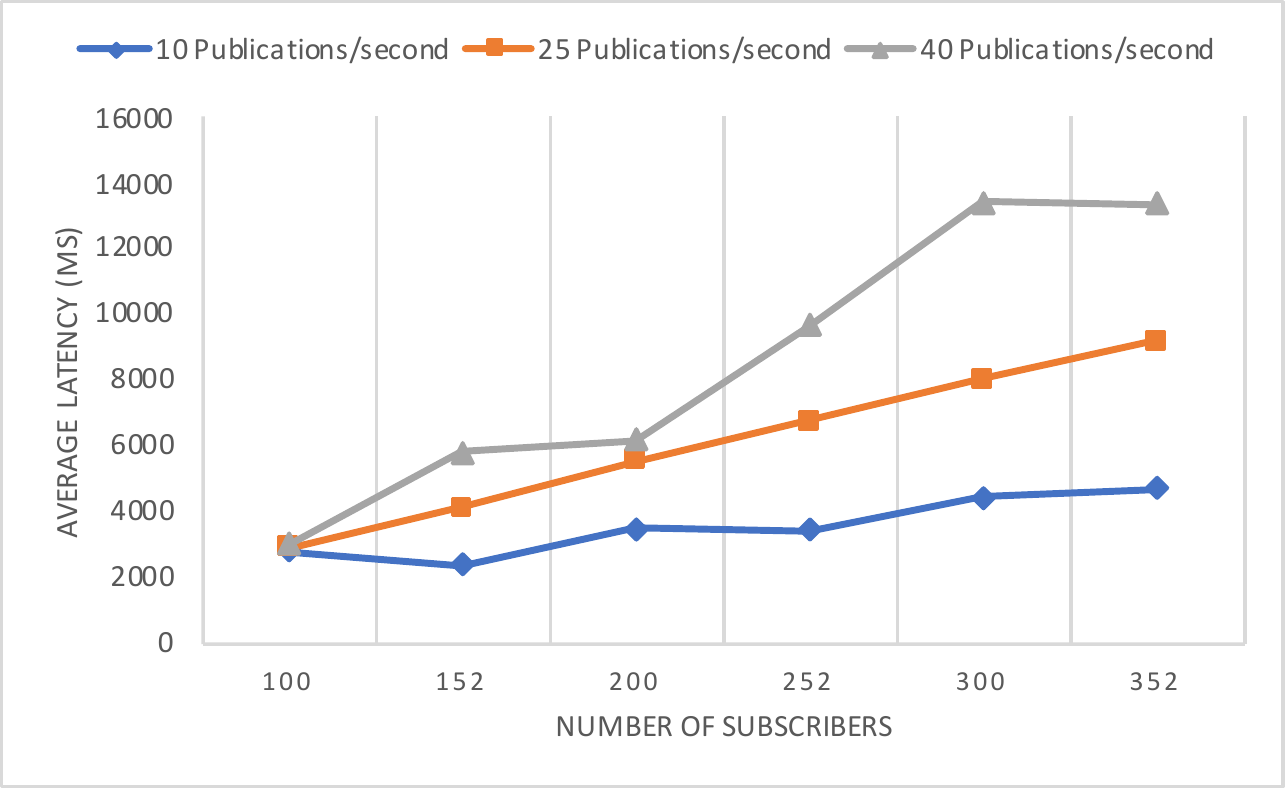}
  \caption{Function-based matching.}
  \end{subfigure}
\caption{The average latency of delivering all the matching publications to the subscribers with a varying number of subscribers and publication throughput. }
\label{fig:evals_result}
\end{figure*}

We summarize the measured latency for the different matching schemes with a varying number of subscribers and publications throughput in Figure \ref{fig:evals_result}. The values on the vertical axis of Figure \ref{fig:evals_result} present the arithmetic mean of latencies for delivering all the submitted publications to the entire set of subscribers in the experiment. The results show that the serverless broker scales up in response to the increasing workload. However, the latency of delivering the publications also increases as the number of subscribers and publications throughput increases. We did not increase the number of subscribers to more than 352 devices nor the number of publications per second rate to more than 40 because above this level. We would hit the IBM Cloud Function limitations of 1200 simultaneous action executions, preventing any new invocations.

\subsection{Discussion and Limitations} 
\label{subsec:disscusion}

Developing a stable, scalable, fault-tolerant pub/sub system requires overwhelming development effort. On a large scale, the developers and system administrators should address the issues raised by coordinating and balancing the load among multiple brokers running in parallel, which increases the cost and effort of maintaining such a system. Nevertheless, since the serverless providers are responsible for maintaining the operational concerns, the development team can focus on creating a pub/sub system rather than the scalability issues. However, our system can perform a wide range of complicated matching schemes and the source code of our serverless consists of less than 1200 lines of JavaScript code. Furthermore, based on our experience, the learning curve for developing serverless applications is relatively short since the actions are programmed based on conventional programming models. The IBM Cloud Functions ecosystem provides a comprehensive set of packages and libraries.
Furthermore, the cost of hosting and executing the serverless applications on the Cloud Functions is calculated based on the actual time the actions are executing and the amount of memory the actions consume during execution. This pricing method can lead to cost reductions compared to traditional server applications hosted on virtual machines since the customers do not need to pay for idle time and underutilized resources. For example, the total cost of performing our evaluations on IBM Cloud Functions was about 4.23 Euro for about 333000 GB-seconds (GB-seconds is the rate Cloud Functions uses for calculating the costs). In summary, we argue that IBM Cloud Functions is a practical tool for the efficient development of elastic server applications.

The primary limitation we face is the stateless nature of the serverless functions. The actions do not persist nor share the application's state for scalability reasons. However, similar to our pub/sub system, this issue severely limits the possible use cases that can be realized. As a workaround, we depend on the Cloudant DB to maintain the application's state. Nevertheless, the Cloudant DB's restrictions on the number of lookups per second bottleneck the application's scalability. Therefore, we had several design iterations to improve and optimize our approach by decreasing the system dependency on the Cloudant DB, such as storing the subscriptions redundantly in the database to avoid multiple lookups and using the local cache in the actions. However, we should emphasize that the serverless actions do not have access to a robust local cache. The Cloud Functions platform wraps the action's code in a container and executes the container in response to the events. The platform can reuse the containers across multiple executions before destroying the containers after being idle for a specific amount of time. Therefore, the cache stored in the container instance lasts as long as the container is kept running. However, the Cloud Functions does not guarantee any period that the instances are running. Therefore, cache misses are frequent and unpredictable and should be used with care.

In addition to the Cloudant DB restrictions, the IBM Cloud Functions also defines some system limits, which are bottlenecks to the system's scalability, including restricting the number of action executions to 1200 concurrent invocations and 9000 invocations per minute. For this reason, during the evaluation, we had to limit the number of participating subscribers to 352 devices.

During the development phase, the only option for the developers to execute the actions is by deploying the actions to the Cloud Functions environment. No tool can simulate, monitor, and debug the functions locally on the developer's machine. The Cloud Functions provides a command-line interface for monitoring the invocations by printing the execution logs and the stack trace in case of failure. However, the records contain limited information, and there is no convenient way to filter the logs based on action or message type. However, the design of our broker is relatively simple and contains only a few actions; monitoring and debugging the application is relatively frustrating. The development becomes more challenging as the number of actions and the chained actions increases. 

The serverless paradigm is a good fit for backend applications, which can be decomposed into short-lived stateless microservices. The serverless paradigm can cut back on development costs and efforts. Although developers can compensate for some of the serverless limitations by integrating other cloud resources, such as databases, the dependence on other resources can affect the application's scalability.

%%%%%%%%%%%%%%%%%%%%%%%%%%%%%%%%%%%%%%%%%%%%%%%
\section{Conclusions} 
\label{sec:conclusion}

In this paper, we presented the design, implementation, and evaluation of a serverless pub/sub system based on the IBM Cloud Functions that perform topic-based and content-based matchings, as well as a novel matching scheme that we call function-based matching. To compensate for the serverless platform limitations, such as the lack of a persistent state, we integrated IBM Cloudant DB. Our evaluations verify that our design scales up in response to the increasing number of subscribers and publications throughput. However, because of the vendor-specific restrictions on the Cloudant DB and Cloud Functions, the database interactions and the limited number of parallel action invocations can be a scalability bottleneck for the serverless broker. \\

In future work, we will implement and deploy our design on other serverless platforms offered by other primary providers such as Amazon, Microsoft, and Google. We will also investigate how our approach scales in comparison and what limitations we will face. Furthermore, we will investigate how our pub/sub systems stand compared to the standard tools such as Kafka, RabbitMQ, and IBM Message Hub. We will also study the alternative methods for persisting state on the Cloud Functions, such as using scalable shared caches. 

\begin{acks}
We sincerely thank Aleksander Slominski, Vinod Muthusamy, and Vatche Ishakian for their valuable input to this work.
\end{acks}

%%%%%%%%%%%%%%%%%%%%%%%%%%%%%%%%%%%%%%%%%%%%%%%%%
\bibliographystyle{ACM-Reference-Format}
\bibliography{bibliography} 

\end{document}